%% file: main.tex
\pdfoutput=1

\documentclass[11pt]{article}

\usepackage[]{acl}

\usepackage{times}
\usepackage{latexsym}

\usepackage[T1]{fontenc}

\usepackage[utf8]{inputenc}
\usepackage{microtype}
\usepackage{inconsolata}
\usepackage{graphicx}
\usepackage{amsmath}
\usepackage{booktabs}
\usepackage{multirow}
\usepackage{adjustbox}
\usepackage{makecell}
\usepackage{mdframed}
\usepackage{microtype}
\usepackage{hyperref}
\usepackage{cleveref}
\usepackage{multirow}
\usepackage{graphicx}
\usepackage{pdfpages}
\usepackage{url}
\usepackage{xcolor}
\usepackage{epsfig}
\usepackage{adjustbox}
\usepackage{amsfonts, amsmath, amssymb}
\usepackage{booktabs} 
\usepackage{comment}
\usepackage{caption, subcaption}
\usepackage{textcomp}
\usepackage{relsize}
\usepackage{stmaryrd}
\usepackage{bbm}
\usepackage{rotating}
\usepackage{helvet}
\usepackage{courier}
\usepackage{natbib}
\usepackage{cleveref}
\usepackage{xspace}
\usepackage{enumitem}
\usepackage{colortbl}
\usepackage{tabularx}
\usepackage{arydshln}
\usepackage{color,soul}

\crefformat{section}{\S#2#1#3}
\crefformat{subsection}{\S#2#1#3}
\crefformat{subsubsection}{\S#2#1#3}
\newcommand\scalemath[2]{\scalebox{#1}{\mbox{\ensuremath{\displaystyle #2}}}}
\newcolumntype{P}{>{\raggedright\arraybackslash}m{0.98\linewidth}}
\newcolumntype{X}{>{\centering\arraybackslash}m{0.05\linewidth}}

\input{dfn}
\title{Taxonomy-guided Semantic Indexing for Academic Paper Search}

\author{
SeongKu Kang$^1$, Yunyi Zhang$^1$, Pengcheng Jiang$^1$, Dongha Lee$^2$, \\\textbf{Jiawei Han}$^1$, \textbf{Hwanjo Yu}$^3$\thanks{Corresponding author}\\
  $^1$University of Illinois at Urbana Champaign \\  $^2$Yonsei University \qquad $^3$Pohang University of Science and Technology \\
  \texttt{\{seongku,yzhan238,pj20,hanj\}@illinois.edu}\\
  \texttt{donalee@yonsei.ac.kr} \quad \texttt{ hwanjoyu@postech.ac.kr}}

\begin{document}
\maketitle

\begin{abstract}
\input{sections/000Abstract}
\end{abstract}

\section{Introduction}
\label{sec:intro}
\input{sections/010Intro}

\section{Related Work}
\label{sec:relwork}
\input{sections/020Related_work}

\input{sections/030Problem_formulation}

\section{\proposed Framework}
\label{sec:method}
\input{sections/040Method}

\section{Experiments}
\label{sec:exp}
\subsection{Experiment setup}
\label{sec:exp_setup}
\input{sections/051Experiment_setup}

\input{sections/052Expriment_result}

\section{Conclusion}
\label{sec:conclusion}
\input{sections/060Conclusion}

\section{Limitations}
\label{sec:limitations}
\input{sections/070Limitation}

\section{Ethical Statement}
\label{sec:ethical_statement}
\input{sections/080Ethical_statement}

\section*{Acknowledgements}
This work was supported IITP grant funded by MSIT (No.2018-0-00584, No.2019-0-01906), NRF grant funded by the MSIT (No.RS-2023-00217286, No.2020R1A2B5B03097210).
It was also in part by US DARPA INCAS Program No. HR0011-21-C0165 and BRIES Program No. HR0011-24-3-0325, National Science Foundation IIS-19-56151, the Molecule Maker Lab Institute: An AI Research Institutes program supported by NSF under Award No. 2019897, and the Institute for Geospatial Understanding through an Integrative Discovery Environment (I-GUIDE) by NSF under Award No. 2118329.

\bibliography{anthology,custom}


\appendix
\input{sections/090Appendix}

\label{sec:appendix}

\end{document}

%% file: dfn.tex
\newcommand{\smallsection}[1]{{\vspace{0.03in} \noindent \bf {#1.\hspace{3pt}}}}

\newcommand{\proposed}{{TaxoIndex}\xspace}

\newcommand{\ctr}{{Contriever-MS}\xspace}
\newcommand{\specter}{{SPECTER-v2}\xspace}
\newcommand{\csfcube}{CSFCube\xspace}
\newcommand{\dorismae}{DORIS-MAE\xspace}

%% file: sections/000Abstract.tex
Academic paper search is an essential task for efficient literature discovery and scientific advancement.
While dense retrieval has advanced various ad-hoc searches, it often struggles to match the underlying academic concepts between queries and documents, which is critical for paper search.
To enable effective academic concept matching for paper search, we propose \textbf{Taxo}nomy-guided semantic \textbf{Index}ing (\proposed) framework. 
\proposed extracts key concepts from papers and organizes them as a \textit{semantic index} guided by an academic taxonomy, and then leverages this index as foundational knowledge to identify academic concepts and link queries and documents.
As a plug-and-play framework, \proposed can be flexibly employed to enhance existing dense retrievers.
Extensive experiments show that \proposed brings significant improvements, even with highly limited training data, and greatly enhances interpretability.

%% file: sections/010Intro.tex
\begin{figure*}[t]
\centering \includegraphics[width=1.0\linewidth]{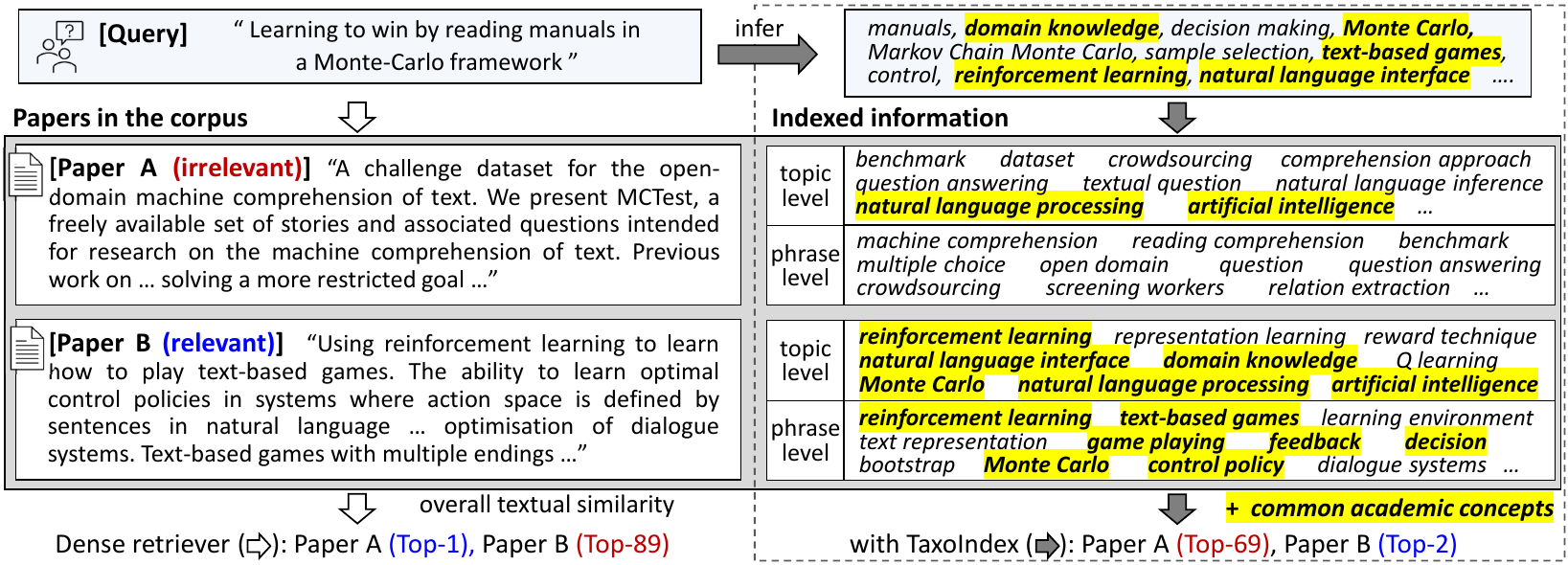}
\caption{A case study from \csfcube dataset. Results of (left) a dense retriever, (right) with \proposed.
For the dense retriever, we use \specter fully fine-tuned on the target corpus. 
}
\label{fig:motivating}
\end{figure*}

Academic paper search is essential for efficient literature discovery and access to technical solutions. 
Recently, dense retrieval has advanced in various ad-hoc searches \cite{karpukhin2020dense, CTR}.
It encodes queries and documents as dense embeddings, measuring relevance by embedding similarity.
These embeddings effectively capture textual meanings via pre-trained language models trained on massive corpora.
While effective in general domains like web search, it often shows limitations in paper search \cite{DORISMAE}.

In paper search, it is crucial to match the underlying academic concepts between queries and documents, rather than relying on surface text and its meanings.
Academic concepts refer to fundamental ideas, theories, and methodologies that constitute the contents of papers. 
Users often seek information on specific concepts when searching for papers.
For example, consider the query "\textit{learning to win by reading manuals in a Monte-Carlo framework}".
This query encompasses various concepts: optimizing decision-making (learning to win), acquiring knowledge from text (reading manuals), and reinforcement learning using probabilistic sampling (Monte-Carlo).
Accordingly, retrievers should find papers that comprehensively cover~these~concepts.

One critical limitation of existing dense retrievers is that such academic concepts are often not effectively captured, making them insufficiently considered in relevance prediction.
Identifying underlying concepts from surface text requires an inherent understanding of domain-specific contents, which is not sufficiently obtained from general corpora.
This challenge is even greater for queries. 
As shown in the previous example, user queries often encompass various academic concepts in highly limited contexts. 
Moreover, queries usually have different expression styles (e.g., terminology choice, language style) from documents, making it difficult to~match~common~concepts.

Figure \ref{fig:motivating}(left) shows the results of a dense retriever for the query, which ranks an irrelevant document (Paper A) at the top-1.
Although the paper mentions `solving a goal' and `comprehension of text', which have similar meanings to `learning to win' and `reading manuals' in the query, its focus is on text comprehension and dataset creation, which largely differs from the query concepts.
This shows that the overall textual similarity captured by dense retrievers is insufficient to reflect specific academic concepts, leading to suboptimal results.

To address this limitation, we introduce a new approach that extracts key concepts from papers in advance, and leverages this knowledge to incorporate academic concepts into relevance predictions.
We construct a \textit{semantic index} that stores semantic components best describing each paper. 
The proposed index represents each paper at two different granularities: topic and phrase levels, as shown in Figure \ref{fig:motivating}(right). 
Topic level provides a broader categorization of research, such as `reinforcement learning' or `natural language processing', while phrase level includes specific details, such as `text-based games' or `control policy', complementarily revealing each paper's concepts.
We leverage this index to enhance the existing dense retrievers, enabling more precise academic concept matching.




We propose \textbf{Taxo}nomy-guided semantic \textbf{Index}ing (\proposed) framework. 
We first introduce a new index construction strategy that extracts key concepts from papers.
To guide this process, we propose using an academic taxonomy, a hierarchical tree structure outlining academic topics.\footnote{Academic taxonomies are widely used for study categorization in various institutions (e.g., \href{https://dl.acm.org/ccs}{ACM Computing Classification System}) and can be readily obtained from the web.}
We then propose a new training strategy, called \textit{index learning}, that trains a model to explicitly identify the indexed information for input text.
This is a critical technique that enables \proposed to infer the most related academic concepts in test queries, even if expressed in different terms or not explicitly mentioned, by associating them with papers having similar contexts.
This inferred information helps find relevant papers sharing academic concepts, combined with textual similarity from dense retrievers.
Figure~\ref{fig:motivating}(right) shows that \proposed finds Paper B as the top-2 result based on high overlap in indexed information.

\proposed offers several advantages for existing dense retrievers.
First, \proposed largely improves retrieval quality by effectively matching academic concepts.
Notably, index learning offers a new type of supervision to identify key concepts from text, which is not directly provided by query-document training pairs, yielding significant improvements even with limited training data.
Second, \proposed enhances interpretability by explicitly representing each text with topics and phrases, as shown in our case studies.
Further, these advancements do not come at a large cost of model complexity. 
\proposed updates only a small module, accounting for 6.7\% of the retriever parameters, yet outperforms~fully~fine-tuned~models.

Our primary contributions are as follows:
(1) We propose \proposed, which systematically constructs and leverages semantic index, for effective academic concept matching in paper search.
(2) We design an index construction strategy that represents each paper at both topic and phrase levels, with the guidance of academic taxonomy.
(3) We introduce an index learning strategy that allows for identifying the most related concepts from an input text.
(4) We evaluate \proposed with extensive quantitative and ablative experiments and~comprehensive~case~studies.

%% file: sections/020Related_work.tex
\begin{figure*}[t]
\centering
\includegraphics[width=1.\linewidth]{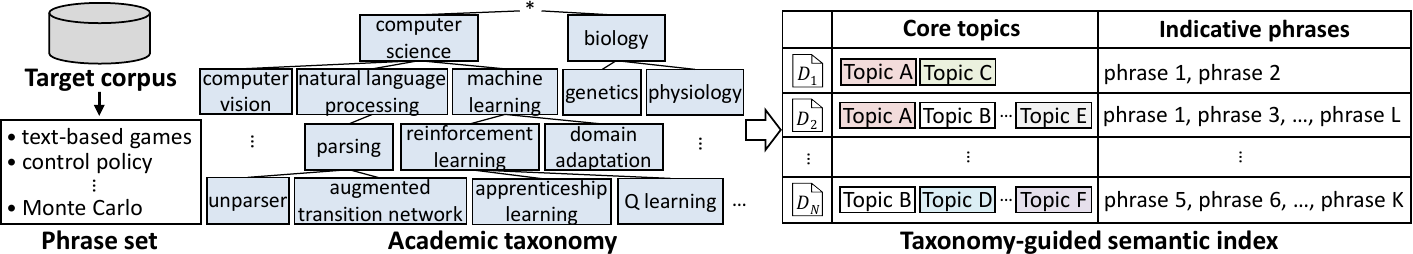}
\caption{A conceptual illustration of the taxonomy-guided semantic index construction. We extract and store core topics and indicative phrases that best represent each paper in the form of a forward index.}
\label{fig:index}
\vspace{-0.04cm}
\end{figure*}



\noindent
\textbf{Dense retrieval.}
The advancement of pre-trained language models (PLMs) has led to significant progress in dense retrieval.
Recent studies have enhanced retrieval quality through retrieval-oriented pre-training \cite{CTR, condenser}, advanced hard negative mining \cite{zhan2021optimizing, rocketqa_v1}, and distillation from cross-encoder \cite{AR2}.
Synthetic query generation has also been explored to supplement training data \cite{BEIR, dai2022promptagator}.
On the other hand, many studies have focused on pre-training methods specialized for the academic domain.
In addition to pre-training on academic corpora \cite{SCIBERT}, researchers have exploited metadata associated with papers. \citet{SPECTER, SCINCL} use citations, \citet{OAGBERT} further utilizes venues, authors, and affiliations. 
\citet{ASPIRE} uses co-citation contexts, and \citet{SPECTER2, zhang2023pre} employs multi-task learning of tasks such as citation prediction~and~paper~classification.

Complementary to the approach of leveraging such paper metadata, we focus on organizing and exploiting knowledge in the textual corpus. 
\proposed can be flexibly integrated to enhance the aforementioned models.

\vspace{0.01in} \noindent
\textbf{Indexing for dense retrieval.}
Indexing refers to the process of collecting, parsing, and storing data to enhance retrieval \cite{indexing}.
Statistical and sparse retrieval often uses inverted indexes for term matching signals \cite{bruch2024efficient}.
Dense retrieval relies on approximate nearest neighbor (ANN) indexes to avoid costly brute-force searches.
Document embeddings are pre-computed offline, and ANN indexes are constructed by techniques such as hashing~\cite{pham2022falconn++}, quantization \cite{baranchuk2018revisiting}, and clustering \cite{zhang2023hybrid, JTR}.

Our index is designed to extract academic concepts and leverage them to improve the accuracy of dense retrievers. 
As \proposed encodes each text as an embedding, existing ANN indexes can still be applied to accelerate search~speed.


%


\vspace{0.01in} \noindent
\textbf{Enhancing retrieval with additional contexts.}
Several studies have enhanced retrieval by providing supplementary contexts. 
Our work falls into this direction.
Pseudo-relevance feedback \cite{BERT-QE, wang2021pseudo, Dense-PRF} utilizes the top-ranked results from an initial retrieval.
Recent generative approaches \cite{mao2021generation, mackie2023generative} generate relevant contexts using PLMs.
\citet{ToTER} utilizes topic distributions of queries and documents.
However, they are often limited in paper search due to the difficulty of generating proper domain-specific contexts.
Moreover, these contexts are obtained and added on-the-fly during inference, making it difficult to provide information tailored~to~backbone~retriever.



%% file: sections/030Problem_formulation.tex
\section{Problem Formulation}
\textbf{Academic taxonomy. }
An academic taxonomy $\mathcal{T}$ refers to a hierarchical tree structure outlining academic topics (Figure \ref{fig:index}).
Each node represents an academic topic, with child nodes corresponding to its sub-topics.
Widely used for study categorization in various institutions, academic taxonomies can be readily obtained from the web and automatically expanded by identifying new topics from a growing corpus \cite{lee2022taxocom, xu2023tacoprompt}. 
We utilize the fields of study taxonomy from Microsoft Academic \cite{MAG_FS}, which covers 19 disciplines (e.g., computer science, biology).

\smallsection{Problem definition}
To perform retrieval on a new corpus $\mathcal{D}$, a PLM-based dense retriever is typically fine-tuned using a training set of relevant query-document pairs.
Our goal is to develop a plug-and-play framework, which facilitates academic concept matching with the guidance of a given taxonomy $\mathcal{T}$, to improve the~backbone~retriever.


%


%% file: sections/040Method.tex
We present taxonomy-guided index construction in \cref{method:s1}, index-grounded fine-tuning in \cref{method:s2}, and retrieval process with \proposed in \cref{method:s3}.

\subsection{Taxonomy-guided Index Construction}
\label{method:s1}
We construct a semantic index that stores semantic components that best describe each paper (Figure \ref{fig:index}).
To guide this process, we propose using the academic taxonomy. 
This ensures that the index organizes knowledge in alignment with the researchers' consensus and greatly improves interpretability.

Our key idea is to represent each paper using a combination of \textit{core topics} and \textit{indicative phrases} that reveal its key concepts at different granularities.
Core topics correspond to nodes in the taxonomy, providing a broader view for categorizing papers. 
Indicative phrases are directly extracted from each paper, offering finer-grained information to distinguish it from other topically similar documents.

%

\subsubsection{Core Topic Identification}
\label{subsub:core_topic}
The given taxonomy may contain many topics not included in the corpus.
To effectively identify core topics from the vast topic hierarchy, we introduce a two-step strategy that first finds candidate topics and then pinpoints the most relevant ones.

\smallsection{Candidate topics identification}
Utilizing the hierarchy, we employ a top-down traversal approach that \textit{recursively visits} the child nodes with the highest similarities at each level.
For each document, we start from the root node and compute its similarity to each child node.
We then visit child nodes with the highest similarities.\footnote{We visit multiple child nodes and create multiple paths, as a document usually covers various topics. For a node at level $l$, we visit $l+2$ nodes to reflect the increasing number of nodes at deeper levels of the taxonomy. The root~node~is~level~$0$.}
This process recurs until every path reaches leaf nodes, and \textit{all visited nodes} are regarded as candidates for the document.

The document-topic similarity ${s}(d, c)$ can be defined in various ways.
As a topic includes its subtopics, we incorporate the information from all subtopics for each topic node.
Let $\mathcal{N}_c$ denote the set of nodes in the sub-tree having $c$ as a root node.
We compute the similarity as: ${s}(d, c) = \frac{1}{|\mathcal{N}_c|}\sum_{j \in \mathcal{N}_c} \operatorname{cos}(\mathbf{e}_{d}, \mathbf{e}_{j})$, 
where $\mathbf{e}_d$ and $\mathbf{e}_j$ denote representations from PLM for a document $d$ and the topic name of node $j$, respectively.\footnote{We use BERT with mean pooling as the simplest choice.}

\smallsection{Core topic selection}
We select core topics by filtering out less relevant ones from the candidates.
We consider two strategies: (1) score-based filtering, which retains topics with similarities above a certain threshold, and (2) LLM-based filtering, which uses large language models (LLMs) to select core topics.
Our preliminary analysis shows that both filtering strategies are effective and lead to comparable retrieval accuracy. 
In this work, we opt for LLM-based filtering, as it often handles ambiguous cases better, further enhancing retrieval interpretability.
Further analysis is provided in \cref{expriment:study}.

We prompt the LLM to select core topics from the candidates by excluding those that are too broad or less relevant.\footnote{The prompt can be found in Appendix \ref{A:prompt}.}
After identifying core topics for all documents, we tailor the taxonomy by only retaining the topics selected as core topics at least once, along with~their~ancestor~nodes.

In sum, for each document $d$, we obtain core topics as $\mathbf{y}^t_d \in \{0,1\}^{|\mathcal{T}_{\mathcal{D}}|}$, where $y^t_{di}=1$ indicates $i$ is a core topic of $d$, otherwise $0$. 
$|\mathcal{T}_{\mathcal{D}}|$ denotes the number of nodes in the tailored taxonomy.



%

\subsubsection{Indicative Phrase Extraction}
\label{method:phrase_identification}
From each document, we extract indicative phrases used to describe its key concepts.
These phrases offer fine-grained details not captured by topic level, playing a crucial role in understanding detailed content and enhancing retrieval.
An indicative phrase should (1) show stronger relevance to the document than to others with similar core topics, and (2) refer to a meaningful and understandable~notion.

We first obtain the phrase set $\mathcal{P}$ in the corpus using an off-the-shelf phrase mining tool \cite{autophrase}.
Then, inspired by \citet{tao2016multi, lee2022taxocom}, we compute the indicativeness of phrase $p$ in document $d$ based on two criteria:
(1) Distinctiveness $dist(p,\mathcal{D}_{d})=\exp(\operatorname{BM25}(p, d))/\,(1 + \sum_{d'\in\mathcal{D}_{d}}\exp(\operatorname{BM25}(p, d')))$ quantifies the relative relevance of $p$ to the document $d$ compared to other topically similar documents $\mathcal{D}_{d}$. 
$\mathcal{D}_{d}$ is simply retrieved using Jaccard similarity of core topic annotation $\mathbf{y}^t_d$.
(2) Integrity $int(p)$ measures the conceptual completeness of the phrase, typically provided by most phrase mining tools, preventing the selection of non-meaningful phrases.
The final indicativeness of $p$ is defined as: $(dist(p,\mathcal{D}_{d}) \cdot int(p))^{\frac{1}{2}}$.

For each document $d$, we select top-$k$ indicative phrases and denote them as $\mathbf{y}^p_d \in \{0,1\}^{|\mathcal{P}|}$, where $y^p_{dj}=1$ indicates $j$ is an indicative phrase of $d$.

\vspace{0.03in} \noindent
\textbf{Remarks.\hspace{3pt}} Compared to recent clustering-based indexes for dense retrieval \cite{zhan2022learning, JTR}, which use cluster memberships from document clustering, the proposed index has several strengths: it effectively exploits domain knowledge from taxonomy, offers broad and detailed views via topics and phrases, and enhances interpretability.





\begin{figure*}[t]
\centering
\includegraphics[width=1.\linewidth]{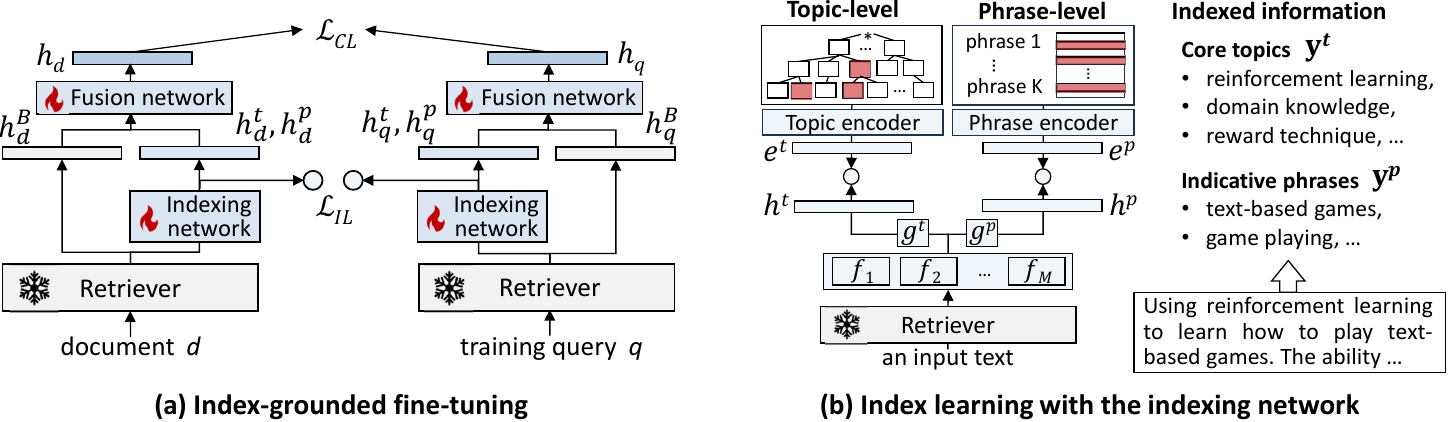}
\caption{An illustration of \proposed: (a) index-grounded fine-tuning, (b) index learning with the indexing network.}
\label{fig:L2I}
\end{figure*}

\subsection{Index-grounded Fine-tuning}
\label{method:s2}
We train an add-on module to enhance relevance prediction while keeping the backbone retriever frozen (Figure \ref{fig:L2I}).
It comprises an indexing network and a fusion network, and is applied identically to both documents and queries using shared parameters.
Here, we describe it~on~the~document~side. 


\subsubsection{Indexing Network: linking text to index}
\label{method:s2a}
A naive approach to using the index information is to append it to each text as additional input context.
However, this approach has several limitations. 
Importantly, test queries are not accessible before the test phase.
Annotating topics and phrases during inference not only incurs additional latency but also is less effective due to the limited~context~of~queries.

As a solution, we propose a new strategy called index learning, which trains the indexing network to identify core topics and indicative phrases from the text.
We formulate this as two-level classification tasks, i.e.,~topic~and~phrase~levels.


%
\smallsection{Extracting topic/phrase information}
Given the backbone retriever embedding $\mathbf{h}_d^B \in \mathbb{R}^l$, we extract information tailored to predict topics and phrases as $\mathbf{h}^t_d$ and $\mathbf{h}^p_d$, respectively.
To exploit the complementarity of topics and phrases, we employ a multi-gate mixture of experts architecture \cite{mmoe}.

We use $M$ different experts, $\{f_m\}_{m=1}^M$, each of which is a small feed-forward network $f_m: \mathbb{R}^l \rightarrow \mathbb{R}^l$.
Two gating networks, $g^t$ and $g^p$, with Softmax outputs control the influence of experts for topic and phrase prediction, respectively.
Let $\mathbf{w}^{t}=g^{t}(\mathbf{h}^B_{d})$ and $\mathbf{w}^{p}=g^{p}(\mathbf{h}^B_{d})$ denote $M$-dimensional vectors controlling the influences. 
The representations for each task are computed as: 
\begin{equation}
    \scalemath{0.94}{\mathbf{h}^{t}_{d} = \sum_{m=1}^{M} w^{t}_{m} f_{m}(\mathbf{h}^B_{d}), \,\, \mathbf{h}^{p}_{d} = \sum_{m=1}^M w^{p}_{m} f_{m}(\mathbf{h}^B_{d})\,}
\end{equation}
This enables the direct sharing of information beneficial for predicting both topics and phrases, mutually enhancing both tasks \cite{mmoe}.

\smallsection{Generating class representation}
We encode topics and phrases to generate class representations for classification learning. 
For topics, we employ graph neural networks (GNNs) \cite{GCN} to exploit their hierarchical information. 
Initially, each node feature is set as the fixed PLM representation of its topic name, followed by GNN propagation over the taxonomy structure.
After stacking GNN layers, we obtain each topic representation $\mathbf{e}^t_i$. 
Each phrase is also encoded using PLM as $\mathbf{e}^p_j$. 
For notational simplicity, we pack the topic and phrase representations, denoting them as $\mathbf{E}^t \in \mathbb{R}^{|\mathcal{T}_{\mathcal{D}}| \times l}$ and $\mathbf{E}^p \in \mathbb{R}^{|\mathcal{P}| \times l}$, respectively.

\smallsection{Index learning}
For each input text, index learning is applied to predict the corresponding core topics and indicative phrases $\mathbf{y}^{t}_{d}$ and $\mathbf{y}^{p}_{d}$.
We compute the probabilities for topics and phrases as $\hat{\mathbf{y}}^{t}_{d} = \operatorname{softmax}(\mathbf{E}^{t} (\mathbf{h}^{t}_{d})^T)$, $\hat{\mathbf{y}}^{p}_{d} = \operatorname{softmax}(\mathbf{E}^{p} (\mathbf{h}^{p}_{d})^T)$, respectively.
The cross-entropy loss is then applied:
\begin{equation}
\scalemath{1.}{
    \mathcal{L}_{IL} = -\sum_{i=1}^{|\mathcal{T}_{\mathcal{D}}|} y^t_{di} \log \hat{y}^{t}_{di} - \sum_{j=1}^{|\mathcal{P}|} y^p_{dj} \log \hat{y}^{p}_{dj}
    }
\end{equation}
The index learning can also be applied to training queries, with topic and phrase labels obtained in two ways: (1) through a separate annotation process, or (2) by using labels of the relevant documents, assuming that they reveal the details of the query.
In this work, we choose the latter approach.


\subsubsection{\scalemath{0.98}{\text{Fusion~Network:~fusing~index~knowledge}}}
The index-based representations ($\mathbf{h}^{t}_{d}, \mathbf{h}^{p}_{d}$) encode core topics and indicative phrases comprising the academic concepts within the text.
We fuse them with the backbone embedding ($\mathbf{h}^{B}_{d}$), which encodes the overall textual meanings, to generate $\mathbf{h}_d \in \mathbb{R}^l$.

We combine the topic and phrase representations as $\mathbf{h}^{I}_d = f^{I}([\mathbf{h}^{t}_{d};\mathbf{h}^{p}_{d}])$ using a small network $f^I: \mathbb{R}^{2l} \rightarrow \mathbb{R}^{l}$.
The final embedding is obtained~as:
\begin{equation}
\label{Eq:fusion}
    \mathbf{h}_d = \mathbf{h}^{B}_d + \alpha \cdot w_d \cdot \mathbf{h}^I_d
\end{equation}
To effectively fuse the two information types, we devise a two-level trainable weight scheme:
(1) a global weight $\alpha$, and (2) an input-adaptive weight $w_d = \operatorname{Sigmoid}(g^I(\mathbf{h}^B_d))$ using a small network $g^I$.
During training, this technique emphasizes index information for input text where predicting relevance from backbone embedding is challenging.


\subsubsection{Fine-tuning with \proposed}
\label{subsubsec:fine-tuning}
We train the add-on module using the standard contrastive learning $\mathcal{L}_{CL}$ with our index learning~$\mathcal{L}_{IL}$. 
For each query $q$, the contrastive~learning~loss~is:
\begin{equation}
    \scalemath{1.}{
    -\log\frac{e^{\operatorname{sim}(\mathbf{h}_q, \mathbf{h}_{d^+})}}{e^{\operatorname{sim}(\mathbf{h}_q, \mathbf{h}_{d^+})} + \sum_{d^-} e^{\operatorname{sim}(\mathbf{h}_q, \mathbf{h}_{d^-})}}
    }
    \label{eq:L_cl}
\end{equation}
where $d^+$ and $d^-$ denote the relevant and irrelevant documents. 
Index learning is applied to both documents and training queries $Q$.
The final objective is $\mathcal{L}_{CL}(Q, \mathcal{D}) + \lambda_{IL}(\mathcal{L}_{IL}(\mathcal{D}) + \mathcal{L}_{IL}(Q))$, where $\lambda_{IL}$ is a hyperparameter to balance the loss.
To ensure $\mathbf{h}^I_d$ contains high-quality information, we initially warm up the indexing network using $\mathcal{L}_{IL}$.




\smallsection{Core topic-aware negative mining}
We devise a new strategy that uses core topics to mine hard-negative documents.
Core topics reveal key concepts based on taxonomy, which may not be effectively captured by the lexical overlap (e.g., BM25) widely used for negative mining \cite{formal2022distillation}.
We utilize both topical and lexical overlaps to select negative documents.
For each $(q, d^+)$ pair, we retrieve $\mathcal{D}_{d^+}$, a set of topically similar documents to $d^+$, using Jaccard similarity of core topics, as done in \cref{method:phrase_identification}.
We then select documents with the highest BM25 scores for $q$ as~negative~samples.

\subsection{Retrieval with \proposed}
\label{method:s3}
Based on the index, \proposed incorporates the similarity of surface texts and the similarity of the most related concepts for relevance prediction. 
This approach enhances the understanding of test queries, enables more precise academic concept matching, and improves paper search.

We introduce advanced inference techniques to further enhance retrieval using topic/phrase predictions ($\hat{\mathbf{y}}^t, \hat{\mathbf{y}}^p$) for queries and documents.

\smallsection{Document filtering based on core topics}
Before applying the retriever, we filter out irrelevant documents that have minimal core topic overlap with the query. 
This step enhances subsequent retrieval by reducing the search space and providing topical overlap information. 
We compute the topical overlap using the inner product of $\hat{\mathbf{y}}^t_q$ and $\hat{\mathbf{y}}^t_d$. 
Documents with low topical overlap are excluded, retaining only the top $x$\% of documents from the entire corpus.
In this work, we set $x=25\%$.
We provide retrieval results with varying $x$ in \cref{expriment:study}.

\smallsection{Interpreting search results}
The topics and phrases with the highest probabilities reveal the academic concepts captured and reflected in relevance prediction.
Comparing query and document predictions allows for interpreting the search results. 
We provide case studies in Figure \ref{fig:motivating} and~Appendix~\ref{A:case_study}.


\smallsection{Expanding query with indicative phrases}
We can expand a query by appending top-$k$ phrases not included in the query. 
The retrieval results using the expanded query~are~denoted~as~\underline{\proposed++}.


%% file: sections/051Experiment_setup.tex
\input{sections/991main_table}

We provide further details on setup in Appendix~\ref{A:setup}.

\noindent
\textbf{Dataset and taxonomy.}
We use two datasets\footnote{We provide results on SCIDOCS dataset in BEIR benchmark \cite{BEIR} in Appendix \ref{A:SCIDOCS}.}: \csfcube \cite{CSFCube} and \dorismae \cite{DORISMAE}, which provide test query collections along with relevance labels on the academic corpus, annotated by human experts and LLMs, respectively.
We use training queries generated by \citet{dai2022promptagator}, as they are not provided in both datasets.
We use the field of study taxonomy from Microsoft Academic \cite{MAG_FS} which contains 431,416 nodes.
After indexing, we obtain 1,164 topics and 3,966 phrases for \csfcube, and 1,498 topics and 6,851 phrases for \dorismae.
For core topic selection in \proposed and baselines that require LLMs, we use \texttt{gpt-3.5-turbo-0125}.


\smallsection{Metrics}
Following \citet{mackie2023generative, ToTER}, we employ Recall@$K$ (R@$K$) for a large retrieval size ($K$), and NDCG@$K$ (N@$K$) and MAP@$K$ (M@K) for a smaller $K$ ($\leq 10$).


\smallsection{Backbone retrievers}
We employ two representative models: (1) \specter \cite{SPECTER2} is a highly competitive model trained using metadata of scientific papers. 
(2) \ctr \cite{CTR} is a widely used retriever fine-tuned using vast labeled data from general domains.

\smallsection{Baselines}
We compare three types of methods for applying and improving the backbone retriever.

(1) Conventional approaches: 
\textbf{no Fine-Tuning}, \textbf{Full Fine-Tuning (FFT)}, \textbf{add-on module Fine-Tuning (aFT)}.
FFT and aFT follow standard contrastive learning with BM25 negatives.
FFT updates the entire backbone retriever, while aFT only updates an add-on module identical~to~\proposed.\footnote{The module accounts for 6.7\% of the model parameters.}


(2) Enhancing retrieval with additional context: 
\textbf{GRF} \cite{mackie2023generative} generates relevant contexts by LLMs. We generate both topics and keywords for a fair comparison.
\textbf{ToTER} \cite{ToTER} uses the similarity of topic distributions between queries and documents, with topics provided by the taxonomy.
We apply both methods~to~FFT.

(3) Enhancing fine-tuning using an index: 
\textbf{JTR} \cite{JTR} constructs a tree-based index via clustering, then jointly optimizes the index and text encoder. 
Though focused on efficiency, it also enhances accuracy with index-based learning.
We impose no latency constraints for a fair comparison.

We provide details on hyperparameters and implementation in Appendix~\ref{A:imp_detail}.

%% file: sections/991main_table.tex
\begin{table*}[t]
\centering
\resizebox{1.\linewidth}{!}{%
\begin{tabular}{cl llllll  llllll} \toprule
& & \multicolumn{6}{c}{CSFCube} & \multicolumn{6}{c}{DORIS-MAE} \\ \cmidrule(lr){3-8}\cmidrule(lr){9-14}
& & N@5 & N@10 & M@5 & M@10 & R@50 & R@100 & N@5 & N@10 & M@5 & M@10 & R@50 & R@100 \\ \midrule
& BM25 & 0.307 & 0.310 & 0.088 & 0.134 & 0.504 & 0.635 & 0.354 & 0.330 & 0.079 & 0.107 & 0.490 & 0.669 \\ \midrule
\parbox[t]{2mm}{\multirow{8}{*}{\rotatebox[origin=c]{90}{SPECTER-v2}}} 
& no Fine-Tuning & 0.352 & 0.337 & 0.108 & 0.151 & 0.524 & 0.680 & 0.385 & 0.360 & 0.079 & 0.113 & 0.551 & 0.709 \\
& FFT & 0.372 & 0.368 & 0.123 & 0.169 & 0.576 & 0.692 & 0.408 & 0.387 & 0.084 & 0.122 & 0.562 & 0.736 \\
& aFT & 0.378 & 0.344 & 0.119 & 0.160 & 0.578 & 0.696 & 0.400 & 0.372 & 0.080 & 0.115 & 0.558 & 0.714 \\ \cmidrule(lr){2-14}
& FFT w/ GRF & 0.331 & 0.317 & 0.112 & 0.152 & 0.561 & 0.705 & 0.400 & 0.379 & 0.087 & 0.123 & \textbf{0.586} & \textbf{0.756} \\
& FFT w/ ToTER & 0.406 & 0.375 & 0.135 & 0.179 & 0.591 & 0.710 & 0.423 & 0.394 & 0.091 & 0.128 & 0.563 & 0.736 \\ \cmidrule(lr){2-14}
& JTR & 0.379 & 0.352 & 0.118 & 0.157 & 0.598 & 0.699 & 0.395 & 0.380 & 0.080 & 0.118 & 0.548 & 0.713 \\
& \cellcolor{gray!15}\proposed & \cellcolor{gray!15}\underline{0.458}$^{\dagger *}$ & \cellcolor{gray!15}\underline{0.417}$^{\dagger *}$ & \cellcolor{gray!15}\underline{0.144}$^{\dagger *}$ & \cellcolor{gray!15}\underline{0.198}$^{\dagger *}$ & \cellcolor{gray!15}\textbf{0.633}$^{\dagger *}$ & \cellcolor{gray!15}\underline{0.741}$^{\dagger *}$  & \cellcolor{gray!15}\underline{0.447}$^{\dagger *}$ & \cellcolor{gray!15}\underline{0.421}$^{\dagger *}$ & \cellcolor{gray!15}\underline{0.104}$^{\dagger *}$ & \cellcolor{gray!15}\underline{0.144}$^{\dagger *}$ & \cellcolor{gray!15}0.578$^{\dagger}$ & \cellcolor{gray!15}\textbf{0.756}$^{\dagger}$ \\
& \cellcolor{gray!15}\proposed++ & \cellcolor{gray!15}\textbf{0.469}$^{\dagger *}$ & \cellcolor{gray!15}\textbf{0.426}$^{\dagger *}$ & \cellcolor{gray!15}\textbf{0.158}$^{\dagger *}$ & \cellcolor{gray!15}\textbf{0.209}$^{\dagger *}$ & \cellcolor{gray!15}\underline{0.621}$^{\dagger *}$ & \cellcolor{gray!15}\textbf{0.746}$^{\dagger *}$ & \cellcolor{gray!15}\textbf{0.449}$^{\dagger *}$ & \cellcolor{gray!15}\textbf{0.424}$^{\dagger *}$ & \cellcolor{gray!15}\textbf{0.105}$^{\dagger *}$ & \cellcolor{gray!15}\textbf{0.145}$^{\dagger *}$ & \cellcolor{gray!15}\underline{0.581}$^{\dagger}$ & \cellcolor{gray!15}\underline{0.751}$^{\dagger}$ \\ \midrule
\parbox[t]{2mm}{\multirow{8}{*}{\rotatebox[origin=c]{90}{Contriever-MS}}} 
& no Fine-Tuning & 0.340 & 0.311 & 0.095 & 0.130 & 0.551 & 0.682 & 0.427 & 0.398 & 0.084 & 0.124 & 0.507 & 0.635 \\
& FFT & 0.364 & 0.328 & 0.110 & 0.149 & 0.589 & 0.705 & 0.453 & 0.407 & 0.093 & 0.132 & 0.510 & 0.652 \\
& aFT & 0.346 & 0.328 & 0.104 & 0.151 & 0.578 & 0.711 & 0.439 & 0.402 & 0.089 & 0.128 & 0.507 & 0.648 \\ \cmidrule(lr){2-14}
& FFT w/ GRF & 0.353 & 0.313 & 0.108 & 0.136 & 0.559 & 0.669 & 0.447 & 0.381 & 0.090 & 0.120 & 0.511 & 0.643 \\
& FFT w/ ToTER & 0.375 & \underline{0.353} & 0.121 & 0.169 & \textbf{0.597} & \underline{0.724} & 0.458 & \textbf{0.423} & 0.097 & 0.139 & 0.539 & 0.703 \\ \cmidrule(lr){2-14}
& JTR & 0.351 & 0.331 & 0.105 & 0.152 & 0.578 & 0.714 & 0.435 & 0.408 & 0.089 & 0.132 & 0.516 & 0.667 \\
& \cellcolor{gray!15}\proposed & \cellcolor{gray!15}\underline{0.400}$^{\dagger *}$ & \cellcolor{gray!15}\textbf{0.386}$^{\dagger *}$ & \cellcolor{gray!15}\underline{0.135}$^{\dagger *}$ & \cellcolor{gray!15}\underline{0.184}$^{\dagger *}$ & \cellcolor{gray!15}\underline{0.596}$^{\dagger}$ & \cellcolor{gray!15}\textbf{0.726}$^{\dagger}$ & \cellcolor{gray!15}\underline{0.461} & \cellcolor{gray!15}\textbf{0.423}$^{\dagger}$ & \cellcolor{gray!15}\underline{0.100} & \cellcolor{gray!15}\underline{0.141}$^{\dagger}$ & \cellcolor{gray!15}\underline{0.557}$^{\dagger *}$ & \cellcolor{gray!15}\underline{0.729}$^{\dagger *}$ \\
& \cellcolor{gray!15}\proposed++ & \cellcolor{gray!15}\textbf{0.421}$^{\dagger *}$ & \cellcolor{gray!15}\textbf{0.386}$^{\dagger *}$ & \cellcolor{gray!15}\textbf{0.144}$^{\dagger *}$ & \cellcolor{gray!15}\textbf{0.185}$^{\dagger *}$ & \cellcolor{gray!15}0.595 & \cellcolor{gray!15}\textbf{0.726}$^{\dagger}$ & \cellcolor{gray!15}\textbf{0.463} & \cellcolor{gray!15}\underline{0.422}$^{\dagger}$ & \cellcolor{gray!15}\textbf{0.101} & \cellcolor{gray!15}\textbf{0.142}$^{\dagger}$ & \cellcolor{gray!15}\textbf{0.560}$^{\dagger *}$ & \cellcolor{gray!15}\textbf{0.733}$^{\dagger *}$ \\
 \bottomrule
\end{tabular}}
\caption{Retrieval performance comparison on CSFCube and DORIS-MAE datasets. $^{\dagger}$ and * indicate the statistically significant difference (paired t-test, $p<0.05$) from FFT and the best baseline, respectively.}
\label{tab:main}
\vspace{-0.3cm}
\end{table*}

%% file: sections/052Expriment_result.tex
\subsection{Retrieval Performance Comparison}
\label{exp:result_a}



\textbf{Main results.}
In Table \ref{tab:main}, \proposed performs better than all baselines on both backbone models across various metrics. 
Notably, \proposed consistently outperforms FFT despite using significantly fewer trainable parameters, and aFT despite using the same add-on module.
This shows the efficacy of the proposed approach using the semantic index.
Conversely, GRF often degrades performance. 
The LLM-generate contexts are not tailored to target documents, potentially causing discrepancies in expressions and focused aspects.\footnote{For example, for the query in Figure \ref{fig:motivating}, `simulation-based learning' is included in the generated topics, which may be related but not covered by papers in the corpus.}
JTR also fails to outperform FFT.
It relies on document clustering, which may be less effective in specialized domains.
Among the baselines, ToTER shows competitive performance by leveraging topic information. 
However, it cannot consider fine-grained concepts not covered by topics, and adds topic information on-the-fly only at inference, failing to fully enhance the backbone retriever.
Lastly, while \proposed++ brings improvements, they are not significantly high, possibly because the phrase information is~already~reflected~by~\proposed.

For the subsequent analyses, we use \specter for \csfcube and \ctr for \dorismae which show the highest NDCGs in~Table~\ref{tab:main}.


\begin{table}[t]
\centering
\resizebox{1\linewidth}{!}{%
\begin{tabular}{cllllll} \toprule
\multirow{2}{*}{\parbox{1.2cm}{\centering Training\\ data}} 
 & \multirow{2}{*}{} & \multicolumn{2}{c}{\csfcube} & \multicolumn{2}{c}{\dorismae} \\  \cmidrule(lr){3-4} \cmidrule(lr){5-7}
\multicolumn{1}{l}{\textbf{}} &  & N@5 & R@50 & N@5  & R@50 \\ \midrule
\multirow{2}{*}{100\%} & FFT$\,$ & +5.92\%  & +9.84\% & +6.09\% & +0.59\% \\
 & \proposed & +\textbf{30.36}\%$^\dagger$  & +\textbf{20.73}\%$^\dagger$ & +\textbf{7.91}\%$^\dagger$  & +\textbf{9.74}\%$^\dagger$ \\ \midrule
\multirow{2}{*}{50\%} & FFT & +0.80\%  & +7.91\% & +3.46\% & \textcolor{red}{- 0.12\%} \\
 & \proposed & +\textbf{20.65}\%$^\dagger\,$  & +\textbf{19.26}\%$^\dagger\,$ & +\textbf{7.58}\%$^\dagger\,$ & +\textbf{8.40}\%$^\dagger$ \\ \midrule
\multirow{2}{*}{10\%} & FFT & +0.51\%  & +0.40\% & +2.29\% &  \textcolor{red}{- 0.41\%} \\
 & \proposed & +\textbf{19.63}\%$^\dagger$ & +\textbf{15.64}\%$^\dagger$ & +\textbf{6.76}\%$^\dagger$  & +\textbf{7.69}\%$^\dagger$ \\ \bottomrule
\end{tabular}}
\caption{Results with varying amounts of training data. 
We report improvements over no Fine-Tuning. 
$^\dagger$ denotes $p < 0.05$ from paired t-test with FFT. }
\label{tab:limited_data}
\end{table}

\begin{table}[t]
\centering
\resizebox{1.\linewidth}{!}{%
\begin{tabular}{clllll} \toprule
&  & \multicolumn{2}{l}{(a) High lexical mismatch} &  \multicolumn{2}{l}{(b) High concept diversity} \\  \cmidrule(lr){3-4} \cmidrule(lr){5-6}
&  & N@5   & R@50   & N@5    & R@50   \\ \midrule
\multirow{3}{*}{\rotatebox[origin=c]{90}{\small CSFCube}} & FFT & +5.81\%& \textcolor{red}{- 1.45\%}&	+5.01\%&		+0.78\% \\
& ToTER & +16.19\%&+11.86\%&+5.10\%&	+10.90\% \\
& \proposed & +\textbf{56.35}\%$^*$&+\textbf{16.94}\%$^*$&+\textbf{50.22}\%$^*$&+\textbf{11.25}\% \\ \midrule
\multirow{3}{*}{\rotatebox[origin=c]{90}{\small DORIS-M.}} & FFT   & +1.04\%&		+3.47\%&	+0.10\%&		\textcolor{red}{- 2.21\%} \\
& ToTER &+1.07\%&		+14.24\%&	+6.90\%&	+4.78\% \\
& \proposed   & +\textbf{8.58}\%$^*$&		+\textbf{24.65}\%$^*$&	+\textbf{10.41}\%$^*$& 	+\textbf{11.16}\%$^*$ \\ \bottomrule
\end{tabular}}
\caption{Further analysis for \textit{difficult queries}. We report improvement over no Fine-Tuning. $^*$ denotes $p < 0.05$ from paired t-test with ToTER.}
\label{tab:difficult}
\vspace{-0.19cm}
\end{table}

\vspace{0.03in}
\noindent
\textbf{Impacts of training data.}
Table \ref{tab:limited_data} reports the improvements by FFT and \proposed with limited training data.
FFT shows restricted improvements with limited data and even degrades performance.
In contrast, \proposed consistently achieves significant improvements, even with highly limited training data.
This is practically advantageous for real-world applications where collecting ample data is challenging.
\proposed trains the model to explicitly identify the most important concepts based on the index knowledge, effectively reducing reliance on the~training~data.


%


\smallsection{Difficult query analysis}
In Table \ref{tab:difficult}, we further analyze results for \textit{difficult queries}, which account for 20\% of total test queries.
They are identified by two factors complicating query comprehension:
(a) high lexical mismatch with documents, (b) high concept diversity within the query.\footnote{We select queries with (a) the lowest BM25 scores for the top-5\% retrieved documents, (b) the highest average pair-wise distance among core topic embeddings of~relevant~documents.}
FFT shows limited results and even degrades performance, despite the overall improvement in Table \ref{tab:main}. 
Conversely, \proposed consistently improves the retrieval quality on both types of queries, effectively handling lexical mismatch and various academic concepts.
These results in \cref{exp:result_a} collectively show the effectiveness of \proposed in academic paper search.



\subsection{Study of \proposed}
\label{expriment:study}

\smallsection{Ablation study}
Table \ref{tab:ablation} presents various ablation results.
First, the best performance is achieved by indexing both topics and phrases.
Notably, removing phrase information drastically degrades performance, as phrases enable fine-grained distinctions of each document.
Conversely, the absence of topic-level can be partially compensated by phrases, leading to smaller performance drops.
Second, both architecture choices improve the indexing network, verifying the efficacy of leveraging the complementarity of topics and phrases and the topic hierarchy.
Lastly, both adaptive weight and topic-aware mining prove effective.
The mining technique shows higher impacts on top-ranked documents and often leads~to~faster~convergence~in~our~experiments.

\begin{table}[t]
\centering
\resizebox{1.\linewidth}{!}{%
\begin{tabular}{lccc}\toprule
 & N@5 & N@10 & R@50\\ \midrule
\proposed & \textbf{0.458} & \textbf{0.417} & \textbf{0.633} \\ \midrule
\multicolumn{4}{l}{\textbf{Indexed information}} \\
\,w/o   Topic-level & 0.415 & 0.382 & 0.619 \\
\,w/o   Phrase-level & 0.385 & 0.369 & 0.578 \\ \midrule
\multicolumn{4}{l}{\textbf{Indexing network architecture}} \\
\,w/o Multi-gate mixture of experts & 0.415 & 0.380 & 0.626 \\
\,w/o GNN & 0.420 & 0.402 & 0.620 \\  \midrule
\multicolumn{4}{l}{\textbf{Training technique}} \\ 
\,w/o Input-adaptive weight (Eq.\ref{Eq:fusion}) & 0.430 & 0.404 & 0.631\\
\,w/o Topic-aware negative mining & 0.432 & 0.403 & 0.630\\ \midrule
aFT (w/o index learning) & 0.378 & 0.344& 0.578 \\
\bottomrule
\end{tabular}}
\caption{Ablation study on \csfcube.}
\label{tab:ablation}
\vspace{-0.1cm}
\end{table}

\begin{figure}[t]
\centering
    \includegraphics[width=0.49\linewidth]{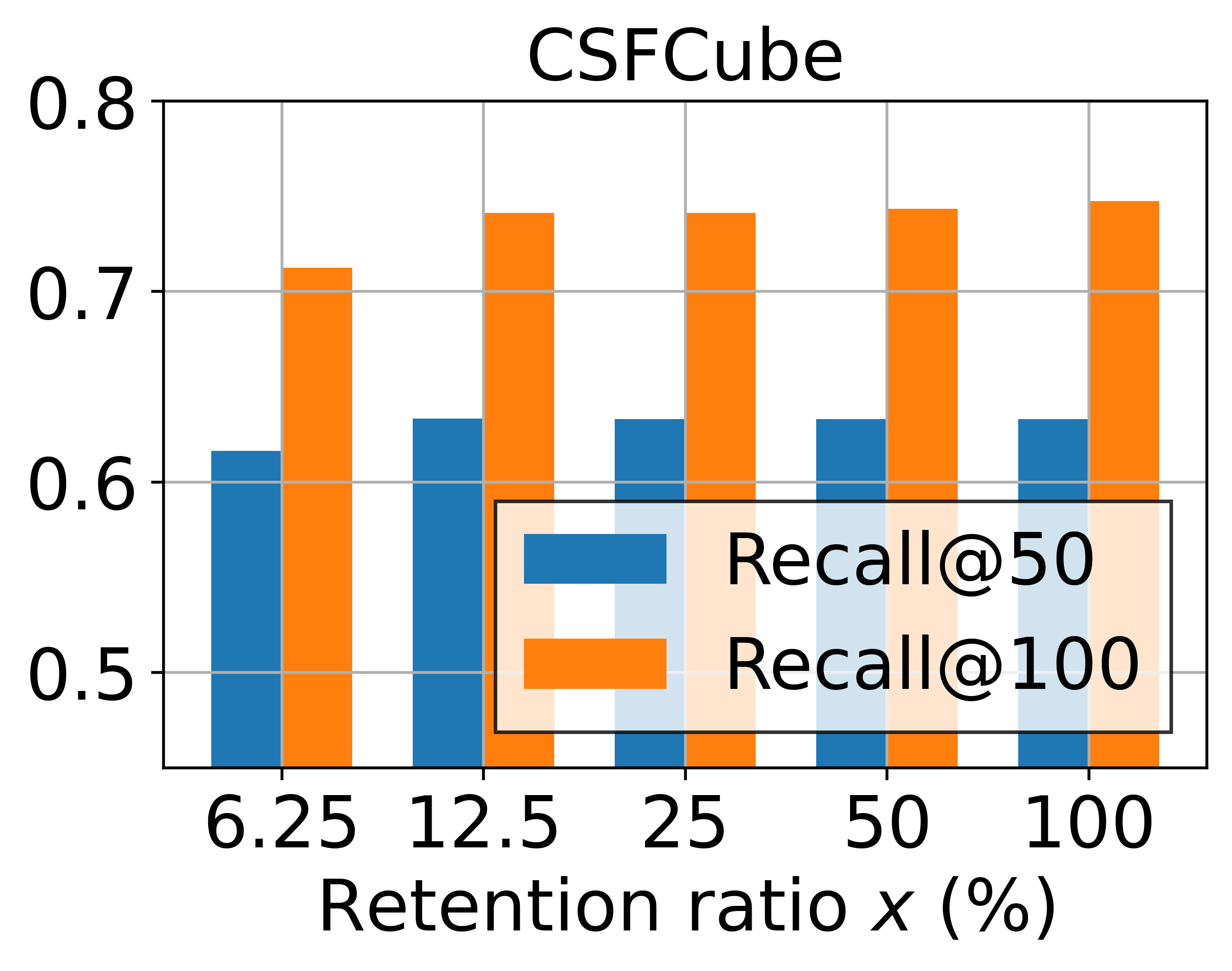}
    \hspace{-0.2cm}
    \includegraphics[width=0.49\linewidth]{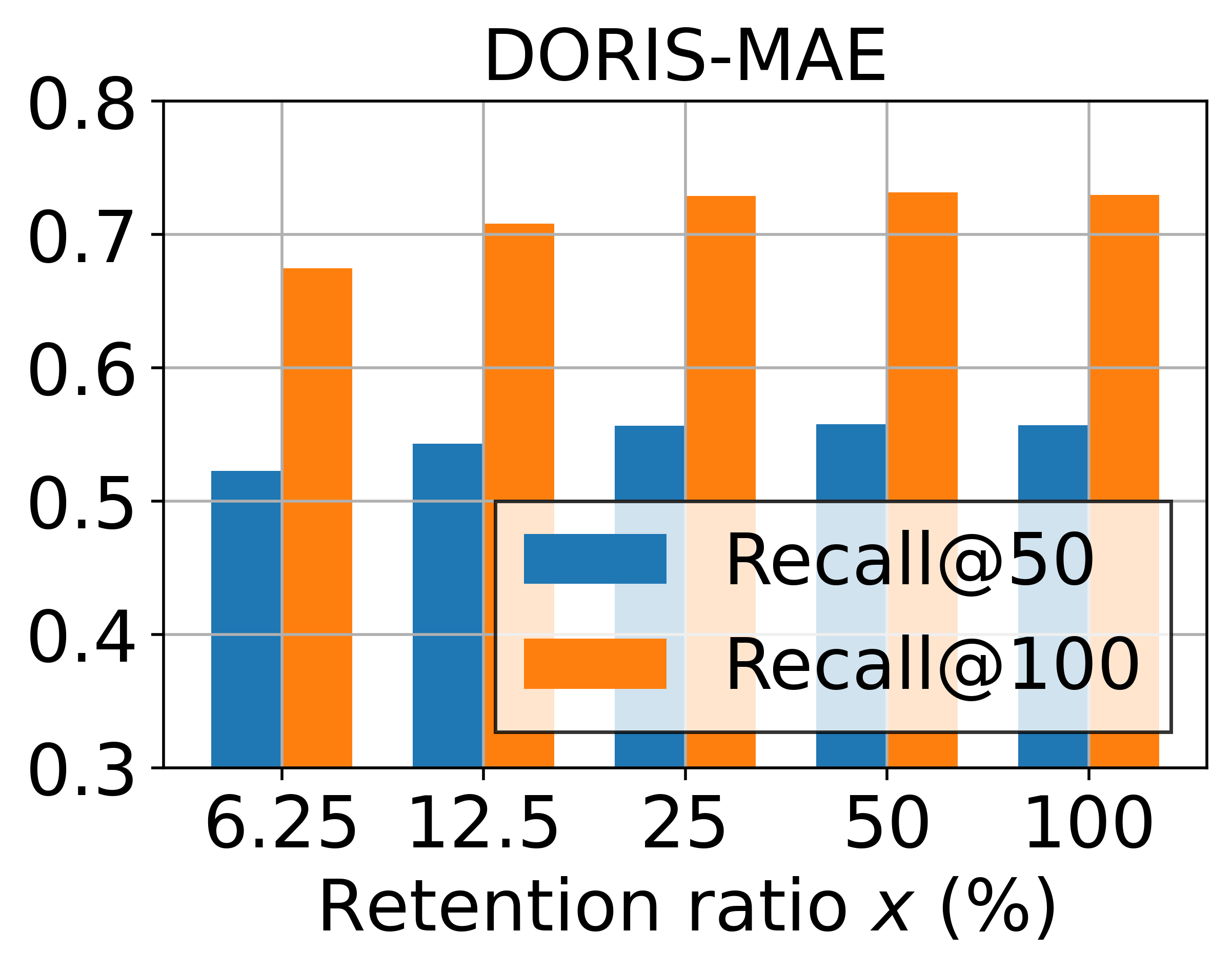}\hspace{-0.1cm}
    \caption{Results with varying retention ratio $x$.}
    \label{fig:retention}
\end{figure}

\smallsection{Document Filtering based on Core Topics}
Figure \ref{fig:retention} shows the retrieval performance with varying retention ratios.
We observe that topic-based filtering achieves comparable results to a whole corpus search by examining about 25\% of the documents.
This result indicates that core topics indeed effectively capture the central theme of each document.

We expect that core topics can be leveraged to improve recent clustering-based ANN indexes \cite{zhan2022learning, JTR}, which conduct clustering on document embeddings and use cluster memberships to represent documents.
As topics are already discrete categories, this approach can reduce the need for clustering operations and provide guidance during the clustering process.
Additionally, it offers interpretability by explicitly using topic names.
As this is not the focus of our work, we leave further investigation for future research.

\begin{figure}[t]
\centering
    \includegraphics[width=0.49\linewidth]{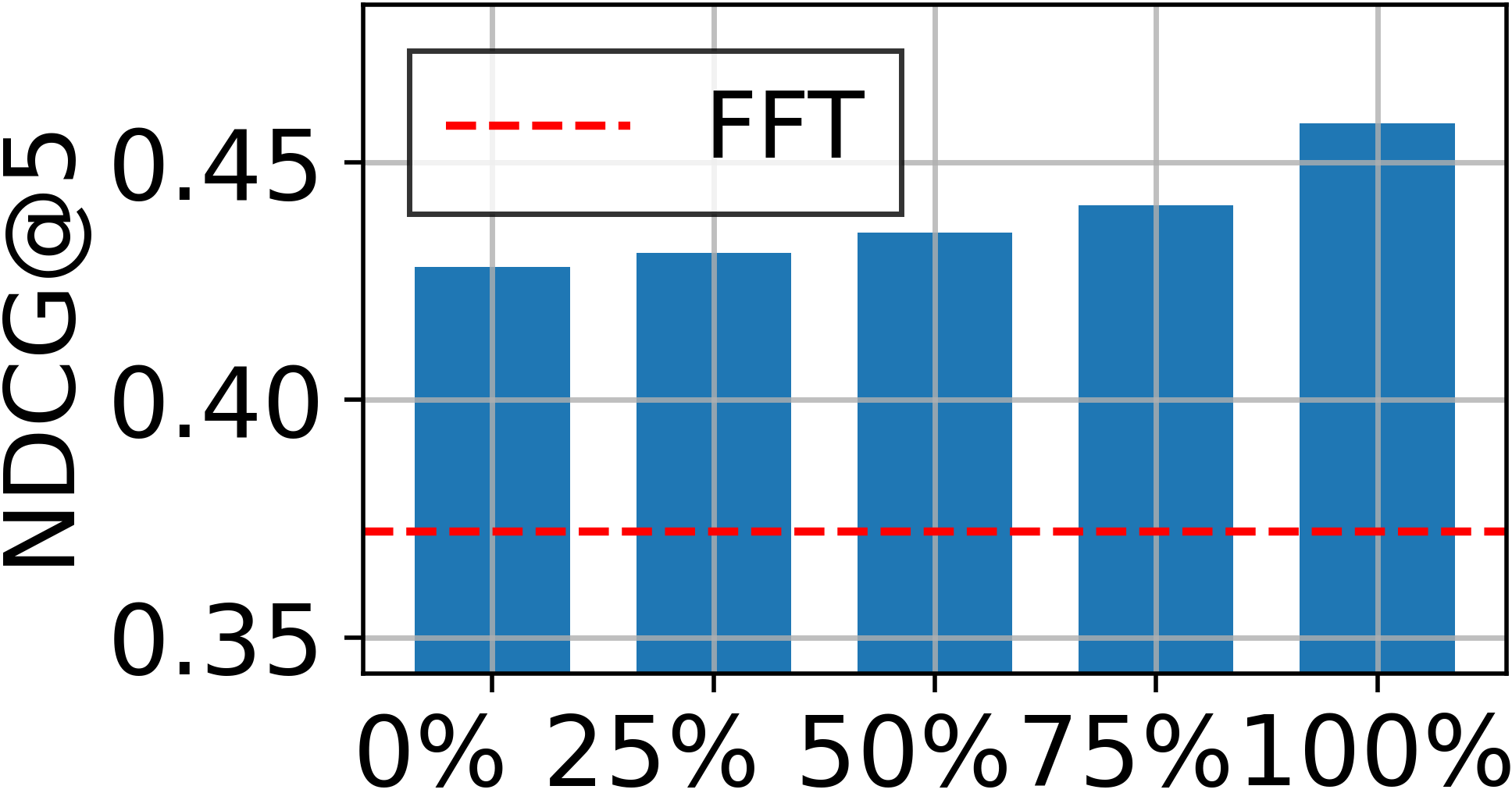}
    \hspace{-0.1cm}
    \includegraphics[width=0.49\linewidth]{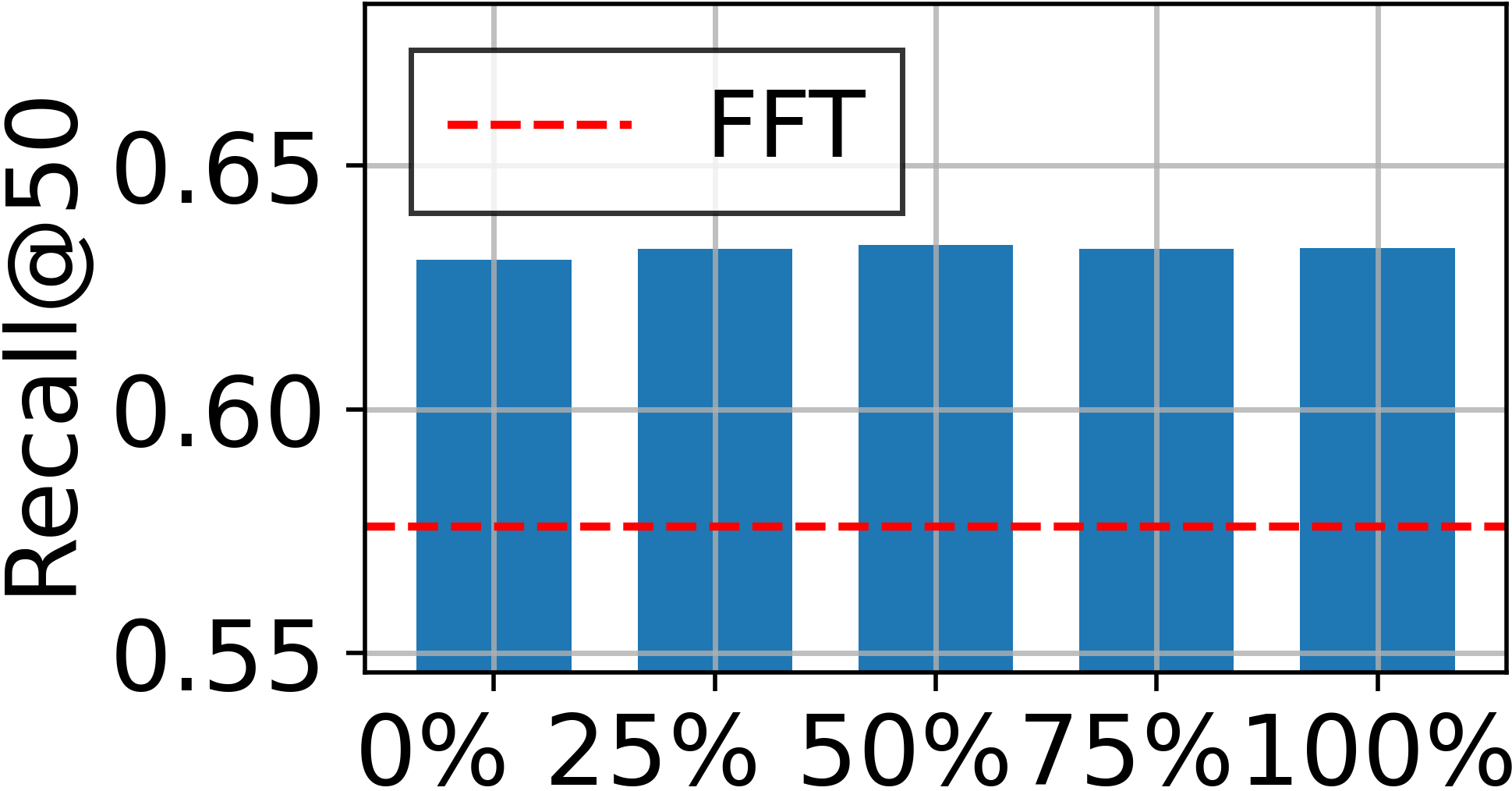}\hspace{-0.1cm}
    \vspace{-0.2cm}
    \caption*{\small (a) Results with varying LLM-based topic filtering ratios.}
    \vspace{0.1cm}
    \includegraphics[width=0.49\linewidth]{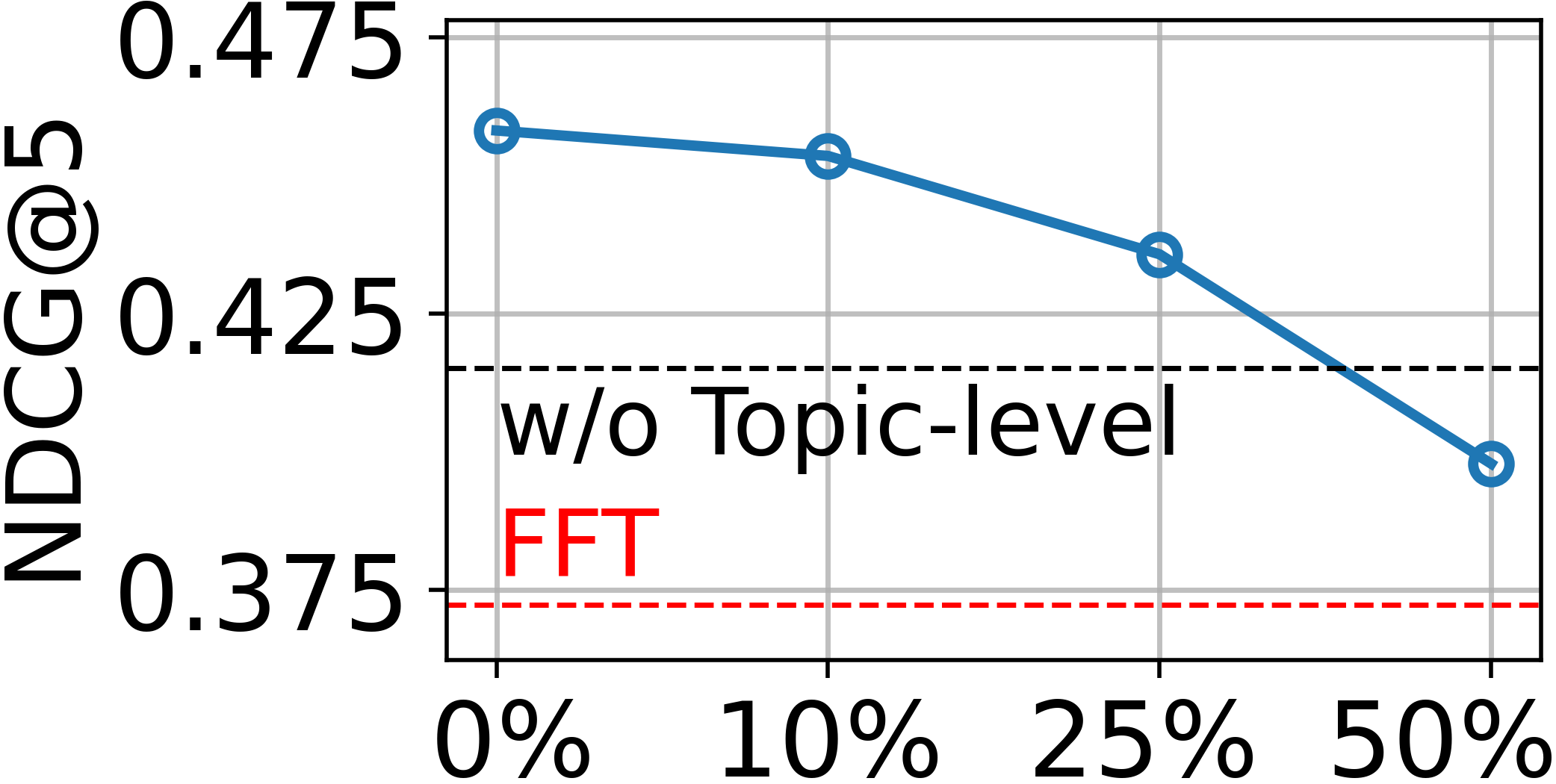}
    \hspace{-0.1cm}
    \includegraphics[width=0.49\linewidth]{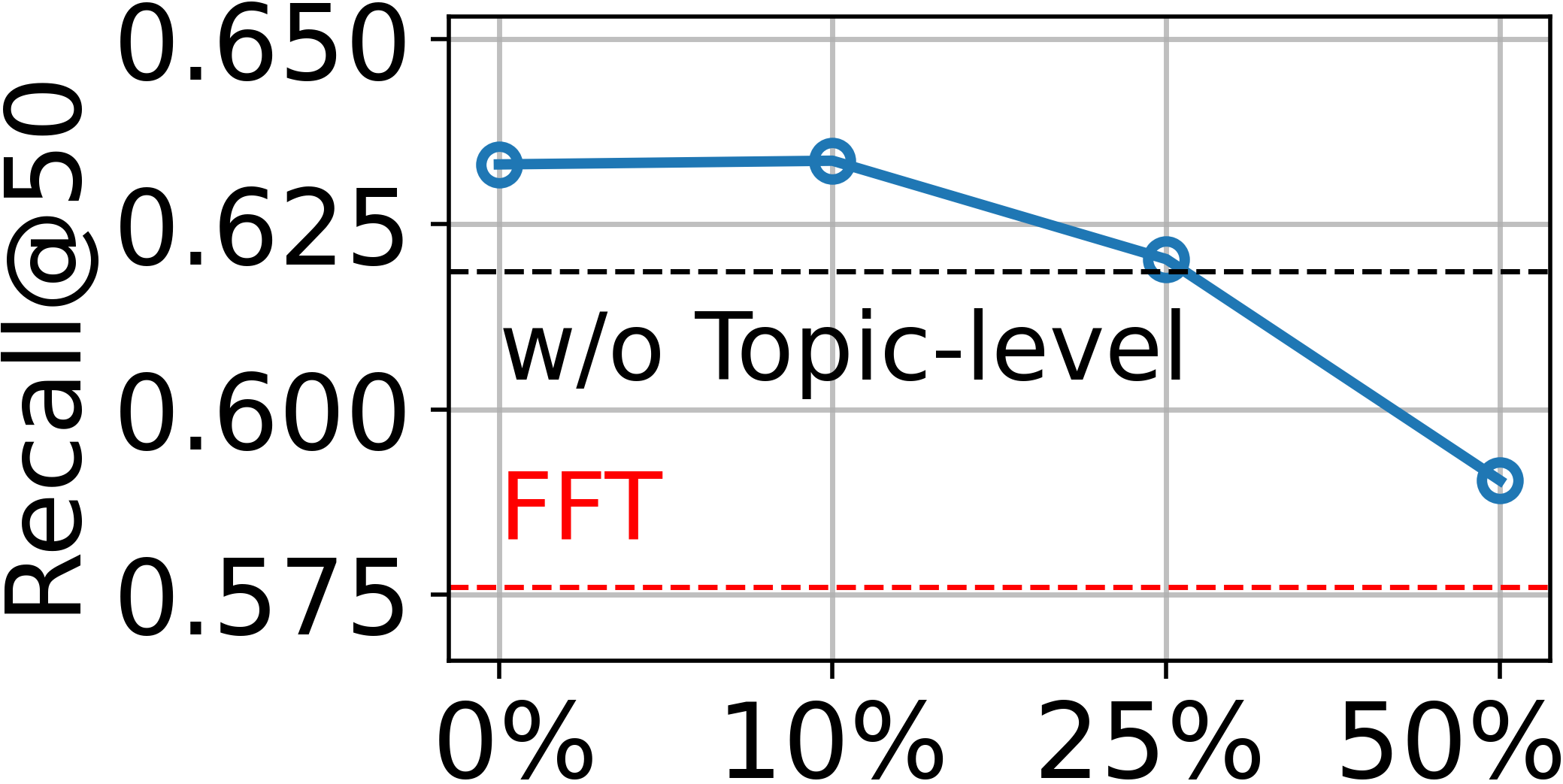}\hspace{-0.1cm}
    \vspace{-0.2cm}
    \caption*{\small (b) Results with varying taxonomy impairment ratios.}
    \vspace{-0.2cm}
    \caption{Taxonomy-related analysis on \csfcube.}
    \label{fig:taxo_analysis}
\end{figure}

\smallsection{Impact of LLM-based topic filtering}
For core topic identification, \proposed utilizes LLM-based filtering (\cref{subsub:core_topic}).
In Figure \ref{fig:taxo_analysis}(a), we explore its impacts by replacing it with score-based filtering, which retains documents with similarity above the median similarity of all documents assigned to each topic.
Both score- and LLM-based filtering consistently achieve significant improvements over FFT.
LLM-based filtering discerns detailed topics better, aiding in finding more precise order among the top-ranked documents, and further enhancing interpretability.
This result also supports the effectiveness of our candidate topic generation strategy.

\smallsection{Impact of taxonomy quality}
\proposed utilizes an academic taxonomy to guide index construction.
In Fig.\ref{fig:taxo_analysis}(b), we explore the impact of taxonomy quality by impairing completeness through random pruning, controlling the ratio of removed nodes to total nodes.
\proposed shows considerable robustness, outperforming FFT even with 50\% pruning.
During training, missing topics can be partially inferred from existing topics and phrases, 
and detailed phrase information can compensate for incompleteness not covered by the topics.
This analysis shows that \proposed is not heavily dependent on taxonomy quality.
We expect \proposed to be effective with taxonomies available on the web, and further improved with existing taxonomy completion techniques \cite{lee2022taxocom, xu2023tacoprompt, zhang2024teleclass}.


%% file: sections/060Conclusion.tex
We propose \proposed to match academic concepts in paper search effectively. 
\proposed extracts key concepts from papers and constructs a semantic index guided by an academic taxonomy.
It then trains an add-on module to identify and incorporate these concepts, enhancing dense retrievers.
Extensive experiments show that \proposed yields significant improvements, even with limited training data.

We expect that \proposed will effectively improve retrieval quality in various domains where underlying search intents are not sufficiently revealed by surface text. 
Specifically, e-commerce \cite{SSCDR, DERRD, MvFS} is an interesting and promising domain. 
In this domain, users often express their information needs in various forms rather than searching by the exact product name. 
They might include desired attributes, characteristics, or even specific use cases. 
\proposed can be applied to better capture users’ search intents in such scenarios. 
We leave further investigation as future work.


%% file: sections/070Limitation.tex
Despite the satisfactory performance of \proposed, our study has three limitations.

First, we utilize an academic taxonomy obtained from the web to guide core topic identification (\cref{subsub:core_topic}).
We acknowledge that the taxonomy may not reflect up-to-date information.
However, we are optimistic that this issue can be addressed by leveraging automatic taxonomy construction and completion techniques, a well-established research fields with many readily available tools \cite{lee2022taxocom, xu2023tacoprompt, shi2024taxonomy}.
Also, our analysis in \cref{expriment:study} shows that \proposed has considerable robustness to taxonomy coverage by utilizing phrase information directly extracted from papers.

Second, for topics and phrase mining process (\cref{method:s1}), we employ relatively simple techniques (e.g., distinctiveness and integrity computations) that have proven effective in recent text mining work. 
While these choices show high effectiveness in our experiments, we acknowledge that more sophisticated techniques could be employed.
Our primary contributions lie in representing each paper's concepts at two levels and incorporating them into relevance predictions, rather than in the specific details for obtaining~topics~and~phrases.

Lastly, this work focuses on the typical dense retrieval models that represent each text as a single vector embedding. 
Applying \proposed to multi-vector representation models \cite{colbertv2} may require additional modifications, which have not been explored in this study.

%% file: sections/080Ethical_statement.tex
We utilize widely recognized and publicly available datasets for research purposes.
Our methodologies and findings do not cause harm to any individuals or groups. 
We do not foresee any significant ethical issues arising from our work.

%% file: sections/090Appendix.tex
\appendix

\section{Prompt for Core Topic Selection}
\label{A:prompt}
We instruct LLMs using the prompt provided below. 
Both datasets used in this work contain paper abstracts.
In our experiments, the average number of candidate topics is $28.5$, and the average number of selected topics is $9.4$.
\begin{table}[h]
    \centering
    \resizebox{0.99\linewidth}{!}{
    \begin{tabular}{|P|}
    \hline
        \textbf{Instruction}: You will receive a paper abstract along with a set of candidate topics for the paper. Your task is to select the topics that best align with the core theme of the paper. Exclude topics that are too broad or less relevant. You may list up to 10 topics, using only the topic names in the candidate set. Do not include any explanation.\\
        \textbf{Paper}: [\textsc{Document}],\\
        \textbf{Candidate topic set}: [\textsc{Candidate topics}]
    \\
    \hline 
    \end{tabular}}
\end{table}

It is important to note that representing the vast number of nodes in the taxonomy within a single prompt is infeasible. 
Our two-step strategy, which first identifies candidate topics and then pinpoints core topics, facilitates effective core topic selection.


\section{Details of Experiment Setup}
\label{A:setup}
\subsection{Dataset}
We have surveyed the literature to find retrieval datasets in the academic domain where relevance is labeled by experts (or annotators with high capabilities).
We select two recently published datasets: \csfcube \cite{CSFCube} and \dorismae \cite{DORISMAE}. 
They provide test query collections along with relevance labels, annotated by human experts and LLMs, respectively. 
They also represent two real-world search scenarios: query-by-example and human-written queries.

For both datasets, we conduct retrieval from the entire corpus including all candidate documents. 
\csfcube dataset consists of 50 test queries, with about 120 candidates per query drawn from approximately 800,000 papers in the S2ORC corpus. 
Annotation scores greater than `2' (nearly identical or similar) are treated as relevant. 
We use the title as the query and both the title and abstract for the documents.
\dorismae dataset consists of 100 test queries, with about 100 candidates per query drawn similarly to \csfcube dataset.
For each query, average annotation scores greater than `1' (the document answers some or all key components) are treated as relevant.

Lastly, we provide results of \ctr on SCIDOCS \cite{SPECTER, BEIR} in Appendix \ref{A:SCIDOCS}.
SCIDOCS dataset contains $1,000$ test queries with relevance labels for $25,657$ papers.
Please note that we exclude this dataset from main experiments, as it uses citation relations for relevance labels, which are utilized for the training of \specter.

\subsection{Academic Taxonomy}
We use the field of study from Microsoft Academic \cite{MAG_FS}, which covers 19 disciplines (e.g., computer science, biology).
It contains $431,416$ nodes and $498,734$ edges with a maximum depth of~$4$.
Note that we prune the taxonomy by only retaining topics included in the target corpus, during the indexing process (\cref{subsub:core_topic}).
The number of nodes after the pruning is provided~in~\cref{sec:exp_setup}.

\subsection{Metrics}
Following the previous work \cite{BEIR, mackie2023generative, ToTER}, we employ Recall@$K$ (R@$K$) for a large retrieval size ($K$), and NDCG@$K$ (N@$K$) and MAP@$K$ for a smaller $K$. 
Recall@$K$ measures the proportion of relevant documents retrieved in the top $K$ results, without consideration of the rank of the documents. 
Conversely, NDCG@$K$ and MAP@$K$ directly consider the absolute rank of each relevant document, where a higher value indicates that relevant documents are consistently found~at~higher~ranks.

\subsection{Experiment Details}
\label{A:imp_detail}

\smallsection{Backbone models}
We use publicly available checkpoints: \specter\footnote{\texttt{allenai/specter2\_base}} and Contriever-MS\footnote{\texttt{facebook/contriever-msmarco}}.
\specter is trained via multi-task learning using paper metadata from SCIBERT \cite{SCIBERT}, and \ctr is fine-tuned via massive training queries (MS MARCO) from BERT base uncased \cite{devlin2018bert}.
Both backbone models have about 110 million parameters.

\smallsection{Computational resources and API cost}
We conduct all experiments using 4 NVIDIA RTX A5000 GPUs, 512 GB of memory, and a single Intel Xeon Gold 6226R processor. 
For ChatGPT API usage, we spent \$11.50 on core topic selection in \proposed and \$39.30 on query generation.

\smallsection{Implementation details}
For BM25, we use Elasticsearch.\footnote{\url{https://github.com/elastic/elasticsearch}}
As training queries are not provided in all datasets used in this work, we leverage synthetic queries generated by using PROMPTGATOR \cite{dai2022promptagator}, the state-of-the-art query generation method.\footnote{We use $\texttt{gpt-3.5-turbo-0125}$ for query generator.}
All compared methods are trained using the same queries. We use 10\% of training data as a validation set. 
We share the generated queries for reproducibility.
We report the average performance over five independent~runs.
\begin{itemize}[leftmargin=*] \vspace{-\topsep}
    \item \textbf{Fine-tuning details.}
    FFT and aFT use top-50 hard negatives mined from BM25 for each query, as done in \citet{formal2022distillation}.
    \proposed uses top-50 hard negatives mined using core topics and BM25 scores (\cref{subsubsec:fine-tuning}).
    We use the inner product as a similarity function for relevance prediction.
    The learning rate is set to $1e^{-6}$ for FFT and $1e^{-4}$ for aFT and \proposed, after tuning among $\{1e^{-7}, 1e^{-6}, ..., 1e^{-3}\}$.
    We set the batch size as $128$ and the weight decay as~$1e^{-4}$.
    
    \item \textbf{GRF, ToTER, and JTR.}
    We utilize the official implementation for GRF\footnote{\url{https://drive.google.com/drive/folders/1LWGTvXGatrAbwbDahYkraK-nim2O9eyN}}, ToTER\footnote{\url{https://github.com/SeongKu-Kang/ToTER_WWW24}}, and JTR\footnote{\url{https://github.com/cshaitao/jtr}}.
    For GRF, we generate both topics and keywords using the same LLMs with \proposed for a fair comparison.
    ToTER and \proposed utilize the same given taxonomy.
    For baseline-specific hyperparameters, we closely follow the recommended values in the original papers and implementations.

    \item \textbf{\proposed.}
    \proposed only updates an add-on module that has 7.38 million parameters, which account for about 6.72\% of backbone models.
    For core topic selection using LLM, we set the temperature as $0.2$.
    The phrase set $\mathcal{P}$ is obtained using AutoPhrase.\footnote{\url{https://github.com/shangjingbo1226/AutoPhrase}}
    In \cref{method:phrase_identification}, the number of phrases per document is set as $k=\min(15, P_d \times 0.2)$, where $P_d$ denotes the total number of phrases in the document $d$.
    This simple choice allows for a natural consideration of the document length.
    The average number of indicative phrases is $13.2$ for CSFCube and $13.7$ for DORIS-MAE.
    For \proposed++, we set $k=15$.
    We set the size of topically similar document set $|\mathcal{D}_{d}|=100$.

    For the indexing network, we set the number of experts as $M=3$, each being a two-layer MLP. 
    $g^t, g^p$ are linear layers with Softmax outputs.
    For the topic encoder, we use a two-layer graph convolution network \cite{GCN}. 
    For the fusion network, we use a two-layer MLP for $f^{I}$, and a linear layer for $g^{I}: \mathbb{R}^{l} \rightarrow \mathbb{R}$.
    
    For index learning of training queries, we use the averaged labels of the relevant documents when a query has multiple relevant documents.
    We first warm up the indexing network using $\mathcal{L}_{IL}$ until the training loss converges.
    We~set~$\lambda_{IL}=0.1$. 
\end{itemize}\vspace{-\topsep}


\section{Supplementary Results}

\subsection{SCIDOCS Results}
\label{A:SCIDOCS}
We provide results of \ctr on SCIDOCS.
Please note that we exclude this dataset from the main experiments, as it uses citation relations for relevance labels, which are utilized for the training of \specter. 
We conduct automatic evaluation using LLMs as well as conventional evaluation using relevance labels.

\smallsection{Conventional evaluation}
Table \ref{tab:scidocs_convention} presents the retrieval results.
\proposed shows higher retrieval performance compared to FFT and ToTER, despite using significantly fewer trainable parameters.

\begin{table}[h]
\centering
\resizebox{1.\linewidth}{!}{%
\begin{tabular}{llllll}\toprule
 & N@5 & N@10  & R@20 & R@50 & R@100 \\ \midrule
FFT & \textcolor{red}{- 7.71\%} & \textcolor{red}{- 5.18\%} & \textcolor{red}{- 2.15\%} & +1.41\% & +4.47\% \\
ToTER & +4.22\% & +6.05\%  & +9.37\% & +10.47\% & +14.00\% \\
\proposed & +\textbf{13.29}\% & +\textbf{14.74}\%  & +\textbf{15.15}\% & +\textbf{16.70}\% & +\textbf{18.56}\%\\ \bottomrule
\end{tabular}}
\caption{Results on SCIDOCS dataset. We report relative performance change with respect to the backbone retriever (i.e., no Fine-Tuning).}
\label{tab:scidocs_convention}
\end{table}

\smallsection{Automatic evaluation}
For a more thorough evaluation with explicit consideration of detailed contents, we leverage  ranking ability of LLMs for automatic evaluation \cite{RRADistill, qin2023large}.
We adopt a recent pair-wise ranking technique \cite{qin2023large} that instructs LLMs to compare the relevance of two passages for a given query.
The prompt is provided below.
\begin{table}[h]
    \centering
    \resizebox{0.99\linewidth}{!}{
    \begin{tabular}{|P|}
    \hline
        Given a query [\textsc{Query}], which of the following two documents is more relevant to the query? 
        Document A: [\textsc{Document A}]\\
        Document B: [\textsc{Document B}]\\
        Output Document A or Document B.
    \\
    \hline 
    \end{tabular}}
\end{table}

\begin{figure}[t]
\centering
    \includegraphics[width=0.99\linewidth]{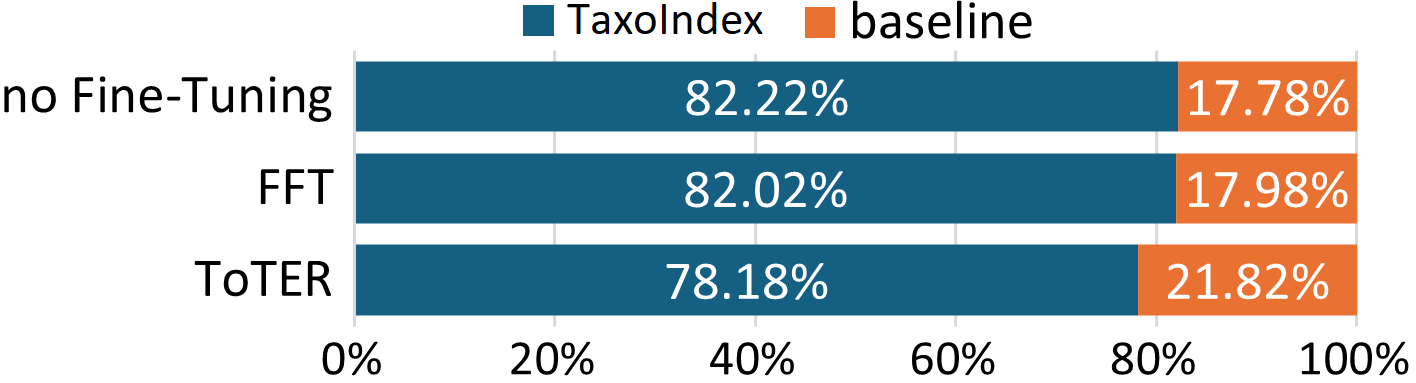}
    \caption{Win ratios of \proposed and each baseline method by automatic evaluation using LLMs. We use $\texttt{gpt-3.5-turbo-0125}$.}
    \label{fig:scidocs_automatic}
\end{figure}

We compare top-1 retrieval results from \proposed and each baseline, applying this evaluation only if the results are not identical. 
Figure \ref{fig:scidocs_automatic} presents the results.
Notably, the improvements revealed through automatic evaluation are much larger compared to those from conventional evaluation.
These results collectively support the effectiveness of \proposed in the academic paper search.

\subsection{Indexing Network Performance}
Table \ref{tab:classification} presents the classification performance of the indexing network.
We report the Precision@10 results on the training set, after the warmup of the indexing network.
We observe high precision for both topics and phrases, indicating they are well captured by the proposed network. 

\begin{table}[h]
\centering
\resizebox{0.7\linewidth}{!}{%
\begin{tabular}{ccc} \toprule
 & \csfcube & \dorismae \\ \midrule
Topic-level & 0.863 & 0.865\\ 
Phrase-level & 0.998& 0.980\\
 \bottomrule            
\end{tabular}}
\caption{Topic and phrase classification performance of the indexing network.}
\label{tab:classification}
\end{table}

\subsection{Additional Case Study}
\label{A:case_study}
\smallsection{Setup}
As discussed in \cref{method:s3}, topics and phrases with the highest predicted probabilities reveal academic concepts captured and reflected for retrieval.
We interpret search results by comparing predictions for queries ($\hat{\mathbf{y}}^t_q, \hat{\mathbf{y}}^p_q$) and documents ($\hat{\mathbf{y}}^t_d, \hat{\mathbf{y}}^p_d$).
In our case studies in Figure~\ref{fig:motivating}, Table~\ref{tab:case_study_query}, and Table~\ref{tab:case_study_DORISMAE}, we use topics and phrases having the highest logit values.
Note that we use $\hat{\mathbf{y}}$ instead of $\mathbf{y}$ for documents, as unlabeled but relevant classes are naturally revealed during training.

\smallsection{Case study: short query with limited context}
In Table~\ref{tab:case_study_query}, we present inferred information for two example queries.
For the query `semantic parsing learning with limited labels', \proposed infers concepts such as `syntactic predicate' and `semi-supervised learning'.
Similarly, for the query `domain adaptation approach for machine translation', \proposed identifies related concepts like `parallel corpus' and `NMT (neural machine translation)'.
This inferred information complements limited query context, facilitating concept matching for paper search.
We highlight that our index is constructed by organizing knowledge in the target corpus, and thus these terminologies are actually used in the papers that users search.

\smallsection{Case study: long and complex query}
In Table~\ref{tab:case_study_DORISMAE}, we explore how \proposed handles long and complex queries by analyzing one that includes various concepts. 
For this query, we present retrieval results: (a) an easy case that is well handled by all baselines (document A), and (b) two difficult cases that are not effectively handled by baselines (documents B and C).

The query encompasses various academic concepts: generative approaches for creating game levels, optimization via reinforcement learning or other differentiable methods, and measuring agent performance. 
Document A is ranked at the top-1 by all compared methods due to its high lexical overlap, directly including terms used in the query (e.g., GAN, generated levels). 
In contrast, documents B and C are not retrieved near the top. 
Unlike document A, they express the concepts using different terms (e.g., neuroevolutionary system), making it difficult to find relevance using surface texts. 
Additionally, document C specifically focuses on surrogate models for a shooter game, which obscures the query concepts like level generation.

\proposed infers the most relevant topics and phrases from the query (highlighted in yellow) and incorporates them into relevance prediction. 
This helps to match the underlying academic concepts, improving retrieval results. 
However, it still shows limited effectiveness for document C, as the overlap of indexed information is relatively small.
We also note that fine-grained aspects of `surrogate model' and `character class' are not fully included in the indexed information, potentially because there are fewer documents covering such concepts in the corpus.
We expect that incorporating other knowledge sources (e.g., knowledge bases) can mitigate these problems. 
We leave further~exploration~for~future~work.

\input{sections/994case_study}

%% file: sections/994case_study.tex
{\renewcommand{\arraystretch}{1.35}
    \begin{table*}[t] \begin{center}
    \small
    \setlength{\tabcolsep}{3pt}
    \begin{tabularx}{\linewidth}{X}
        \toprule
        \textbf{Query} \\
        semantic parsing learning with limited labels  \\
        \hdashline
        \textbf{Inferred core topics and indicative phrases}\\
        parsing, labeled data, statistical parsing, parser combinator, top down parsing, syntactic predicate, text annotation, semi-supervised learning, morphological parsing, natural language processing, artificial intelligence\\
        semantic parsing, semantic parsers, taggers, syntactic parsing, parse tree, semantic role labeling, treebanks, predicates, semantic representations, formal semantics, predicate argument, linguistic knowledge, training labels, ...\\ \midrule
        \textbf{Query} \\
        domain adaptation approach for machine translation\\
        \hdashline
        \textbf{Inferred core topics and indicative phrases}\\
        domain adaptation, machine translation, semantic textual similarity, semantic translation, translation probabilities, weakly supervised learning, neural machine translation, natural language processing, machine learning, artificial intelligence\\
        neural machine translation, statistical machine translation, nmt, smt, cross sentence, adaptation techniques, domain adaptation, parallel sentences, parallel corpus, out of domain, model adaptation, translators, machine translation, ...\\
        \bottomrule
    \end{tabularx}
    \vspace{-3pt}
    \caption{Case study for queries having limited contexts (Corpus: \csfcube)}
    \vspace{-10pt}
    \label{tab:case_study_query}
\end{center}
\end{table*}}

{\renewcommand{\arraystretch}{1.35}
    \begin{table*}[t] \begin{center}
    \small
    \setlength{\tabcolsep}{3pt}
    \begin{tabularx}{\linewidth}{X}
        \toprule
        \textbf{Query} \\
        I am seeking a generative modeling approach capable of creating new levels and potentially game settings/environments for a video game with multiple existing levels of difficulty. Specifically, I am interested in exploring how Generative Adversarial Networks (GANs) and other generative methods could generate entirely new levels by emulating the style of previous ones. It is crucial that the newly generated levels are not merely derivative and that my generative model can optimize specific properties, such as the intensity or graphic nature of the game. Given that these properties are non-differentiable, I need a method to either render them differentiable or employ a reinforcement learning-centric approach to optimize these rewards. After generating a variety of levels, I require a method to select some of the best ones. One potential solution could be to evaluate the generated levels using an automatic metric, such as the performance of an AI agent playing the level. Alternatively, I am considering designing a derivative-free stochastic optimization algorithm to guide the search across the space of all synthetically generated levels, steering towards those that meet specific objectives.
 \\
        \midrule
        \textbf{Document A: an \textit{easy} case (Top-1 by all compared methods)} \\
        Illuminating mario scenes in the latent space of a generative adversarial network. Generative adversarial networks (GANs) are quickly becoming a ubiquitous approach to procedurally generating video game levels. While GAN generated levels are stylistically similar to human-authored examples, human designers often want to explore the generative design space of GANs to extract interesting levels. However, human designers find latent vectors opaque and would rather explore along dimensions the designer specifies, such as number of enemies or obstacles. We propose using state-of-the-art quality diversity algorithms designed to optimize continuous spaces, i.e. MAP-Elites with a directional variation operator and Covariance Matrix Adaptation MAP-Elites, to efficiently explore the latent space of a GAN to extract levels that vary across a set of specified gameplay measures. In the benchmark domain of Super Mario Bros, we demonstrate how designers may specify gameplay measures to our system and extract high-quality (playable) levels with a diverse range of level mechanics, while still maintaining stylistic similarity to human authored examples. An online user study shows how the different mechanics of the automatically generated levels affect subjective ratings of their perceived difficulty and appearance. \\
        \midrule
        \textbf{Document B: a successful \textit{difficult} case (FFT: top-62, ToTER: top-39, \proposed: top-9)} \\
        Co-generation of game levels and game-playing agents. Open-endedness, primarily studied in the context of artificial life, is the ability of systems to generate potentially unbounded ontologies of increasing novelty and complexity. Engineering generative systems displaying at least some degree of this ability is a goal with clear applications to procedural content generation in games. The Paired Open-Ended Trailblazer (POET) algorithm, heretofore explored only in a biped walking domain, is a coevolutionary system that simultaneously generates environments and agents that can solve them. This paper introduces a POET-Inspired Neuroevolutionary System for KreativitY (PINSKY) in games, which co-generates levels for multiple video games and agents that play them. This system leverages the General Video Game Artificial Intelligence (GVGAI) framework to enable co-generation of levels and agents for the 2D Atari-style games Zelda and Solar Fox. Results demonstrate the ability of PINSKY to generate curricula of game levels, opening up a promising new avenue for research at the intersection of procedural content generation and artificial life. At the same time, results in these challenging game domains highlight the limitations of the current algorithm and opportunities for improvement. \\ \hdashline
        \textbf{Indexed information} (\hl{Inferred from the query})\\
        \hl{video game development}, \hl{heuristics},  \hl{game design}, intelligent agent, \hl{genetic algorithm}, \hl{reward technique}, optimization problem, \hl{evolutionary algorithm}, \hl{reinforcement learning}, heuristic evaluation, \hl{generative model}, \hl{machine learning}, \hl{artificial intelligence}\\
        \hl{game playing}, \hl{game levels}, \hl{video games}, game design, \hl{agents}, players, general video game ai, \hl{exploration and exploitation}, GVGAI, \hl{simultaneously learn}, \hl{optimization}, autonomous agents, content generation, knowledge acquisition, \hl{rewards}, …\\
        \midrule
        \textbf{Document C: an unsuccessful \textit{difficult} case (FFT: top-258, ToTER: top-155, \proposed: top-108)} \\
        Pairing character classes in a deathmatch shooter game via a deep-learning surrogate model. This paper introduces a surrogate model of gameplay that learns the mapping between different game facets, and applies it to a generative system which designs new content in one of these facets. Focusing on the shooter game genre, the paper explores how deep learning can help build a model which combines the game level structure and the game's character class parameters as input and the gameplay outcomes as output. The model is trained on a large corpus of game data from simulations with artificial agents in random sets of levels and class parameters. The model is then used to generate classes for specific levels and for a desired game outcome, such as balanced matches of short duration. Findings in this paper show that the system can be expressive and can generate classes for both computer generated and human authored levels.\\ \hdashline
        \textbf{Indexed information} (\hl{Inferred from the query})\\
        simulation, surrogate model, \hl{game design}, game mechanics, modeling and simulation, parameter, \hl{video game development}, intelligent agent, \hl{reinforcement learning}, \hl{generative model}, \hl{machine learning}, \hl{deep learning}, \hl{artificial intelligence} \\
        surrogate model, artificial agents, game data corpus, simulation, \hl{environments}, \hl{game levels}, \hl{game playing}, content generation, \hl{agents}, players, characters, character levels, general video game ai, parameters, game design, \hl{video games}, generators, ...
        \\
        \bottomrule
    \end{tabularx}
    \vspace{-3pt}
    \caption{Case study for a long and complex query. All documents are labeled as relevant (Corpus: \dorismae).}
    \vspace{-10pt}
    \label{tab:case_study_DORISMAE}
\end{center}
\end{table*}}